\documentstyle[aps,prb]{revtex}
\begin{document}
\draft

\twocolumn[\hsize\textwidth\columnwidth\hsize\csname @twocolumnfalse\endcsname

\title{Comment on "Magnus Force  in   Superfluids and Superconductors"
}
\author{  G.E. Volovik}

\address{Low Temperature Laboratory, Helsinki University of
Technology, 02150 Espoo, Finland, \\ and \\ Landau Instute for
Theoretical
Physics,
117334
Moscow, Russia}

\date{\today}
\maketitle

\begin{abstract}
The effective action for the Josephson junction arrays (JJA) should contain
the topological term, which violates the particle-hole symmetry.
This term is responsible for the nonzero Magnus force acting on the
vortex. The Magnus force is however small because the parameter of
the particle-hole asymmetry in superconductors is   of order
$\Delta^2/E_F^2 \ll 1$.

\end{abstract}
\


\

]
\narrowtext

 Sonin \cite{Sonin} recently
presented an extreme view on the Magnus force acting on vortices in the
Josephson junction arrays (JJA). He suggested that the "Hall effect is {\it
exactly} absent in the classical theory of JJA which neglects the charge
quantization. Since the Hall effect is linear in the amplitude of the effective
Magnus force, the latter also vanishes in the classical JJA. This statememnt
directly follows from the symmetry of the dynamic equations." He argued that on
the microscopical level the suggested symmetry of equations "is a direct result
of the  {\it particle-hole symmetry}". Here we discuss why this is not true.

The vortices in the arrays could be
considered as massive particles with long-range Coulomb
interaction \cite{EckernSchmid,FazioSchoen}. In the experiment
\cite{ballistic} the straightforward ballistic motion of vortices was
observed which implies the Magnus force, acting on a vortex
perpendicular to its velocity, is absent or is very small. This
was also confirmed by more recent experiments: vortices move
perpendicular to the driving current
\cite{Lachenmann}, and no Hall effect was detected in the system
\cite{noHall}.

On the macroscopic level the general form of the balance of forces acting on the
vortex in the case when the normal component is pinned by the crystal lattice
or by JJA is the Eq.(23) of Ref.\cite{Sonin}:
\begin{equation}
\rho_M {\bf v}_L\times {\bf{\vec \kappa}} + \eta {\bf v}_L={\bf j}\times
{\bf{\vec \kappa}}.
\label{Magnus}
\end{equation}
Here the first term on the left-hand side of the equation is the definition
of the {\it effective Magnus force} according to \cite{Sonin},  $ {\bf{\vec
\kappa}}$ is the circulation vector and
${\bf v}_L$ is the velocity of the vortex with respect to the array. The
second term on the left-hand side is the friction force which we do not
discuss. The force on the right-hand side is produced by the
electric supercurrent
${\bf j}_e=(e/m){\bf j} $, where  ${\bf j}=\rho_s {\bf v}_s$ is the mass
current.

In the ideal case (see below) the parameter $\rho_M$  equals the superfluid
density $\rho_s$, which at $T=0$ and for the translationally invariant
system equals the total mass density $\rho=mn$. The experiments on JJA show that
$\rho_M$ is either zero or is very small compared to $\rho_s$. There are several
controversial exlpanations of the absence of the Magnus force.

(1) In Ref.\cite{FazioOtterloSchoen,BlanterFazio}  it is assumed
that the phase $\phi_i$ of the condensate in the $i$-th island is
canonically conjugated to the electric charge $Q_i$ of the island.
From this assumption it follows that the Magnus force is proportional
to the offset charges on the superconducting islands, rather than to
$\rho$. The effect of offset charges is negligible, in
particular because the "real samples are usually characterized by random offset
charges. As a result the Magnus force averages to approximately zero". The
drawback of this approach is that because of the separate conservation law for
the number of electrons, one should expect that the phase  $\phi_i$ of the
electron condensate is to be canonically conjugated to the number of the
electrons $N_i$ of the island, rather than to the charge $Q_i$ of the island
which is given by the difference in the numbers of electrons and protons.

(2) The more traditional point of view,  that the
Magnus force is proportional to the density of superconducting
electrons on the islands averaged over distances large compared to the
lattice constant of the array (see eg
\cite{GaitanJJA}), contradicts to the experiment. To match the
experiment it was assumed in \cite{ZhuTanAo} that the force is
proportional to the {\it local} superconducting density at the point
where the vortex is situated. Since the vortex does not move through
superconducting islands but through the junctions, the Magnus force on
the vortex can be substantially reduced. This approach can be applied
to the systems in which the Magnus force can be locally
determined, for example if the order parameter
changes smoothly on the distance of the core size. This is apparently
not the case in the JJA.

(3) In Ref.\cite{MakhlinVolovik} the absence of the Magnus force was
ascribed to the nearly complete cancellation of the Magnus force by
the spectral-flow force: $\rho_M=m(n-n_0)$, where $n_0$ deviates from the
particle density $n$ only due to small particle-hole asymmetry. Such
cancellation is known for the bulk superfluids and superconductors in the so
called hydrodynamic regime, where the momentum exchange between the
electrons in the core and that in the heat bath is maximal. In this case
the spectral flow along the low-energy levels of the bound states in the
core of vortices almost completely
cancels the Magnus force
\cite{KopninVolovik1995,Stone,KopninVolovik1997,Makhlin}. It was shown in
\cite{MakhlinVolovik} that similar spectral flow can take place in the case
when the Josephson junctions are of  the
Superconductor--Normal-metal--Superconductor (SNS) type. However this mechanism
cannot be applied to the Josephson junctions of
Superconductor--Insulator--Superconductor (SIS) type, where the low-energy
levels are absent and the spectral flow is forbidden.

(4) Sonin \cite{Sonin}
suggested that the effective Magnus force in the
JJA is exactly zero due to exact symmetry of the motion equations which follows
from the particle-hole symmetry. However if one continuously increases the
transparency of the contact, ie the critical current through the
Josephson contact, one finally reaches the limit of the bulk superconductivity,
where the Magnus force is too well known to be nonzero, at least in the
nonhydrodynamic regime. So one should assume  that on the way from JJA to bulk
superconductivity, there is a quantum Lifshitz transition from zero to nonzero
value of the Magnus force. This does not seem very unreasonable, but one must
take into account that  the particle-hole symmetry is not
exact in superconductors. There is always a small asymmetry of order
$\Delta^2/E_F^2$, where
$\Delta$ is the superconducting gap and $E_F$ is the Fermi energy. This
asymmetry never disappears and thus the Magnus force is never
exactly zero. Let us discuss how this asymmetry can enter the Magnus
force in JJA.

The Magnus force in the systems without translational invariance
still remains an open question, though this problem arises even in
the bulk superconductor due to the band structure of the
electrons in crystals \cite{Volovik1996}. The main problem is what
electrons are involved in the construction of the transverse force
acting on the vortex: are these the extra electrons due to the
offset charge or all electrons; all  superconducting
electrons, or only the Cooper pairs concentrated in a small belt in
momentum space with dimension of the gap
$\Delta$ in the vicinity of the Fermi surface? It can be also the
Andreev bound states in the vortex core or Bogoliubov quasiparticles in
the normal component outside the core.

There is no unique answer to this question, since the result  depends
on the kinetics of the electrons on the background of the moving
vortex. However there are some limiting cases in which the dissipation
can be neglected and the answer can be guessed from the general
principles before the detailed calculations. For example, if two
conditions are fulfilled -- (i) the system is translationally invariant
and (ii) the transport is adiabatic -- the Magnus force has its maximal
{\it ideal} value determined at $T=0$ by the total electron density:
$\rho_M=mn$. This limit occurs eg. when there is a gap in the spectrum of the
electrons even in the presence of vortex cores. In this case the results of
calculations, made by  groups with essentially different ideology, say
\cite{Thouless} and \cite{KopninVolovik1995}, coincide.

If any of the two conditions are violated the life becomes
more complicated.
If the
translational invariance (condition (i)) is absent, the Magnus force is
to be reduced:  some electrons are pinned by the crystal lattice
and thus cannot contribute the transverse force. The pinned
electrons are either those which belong to the completely occupied
electronic bands\cite{Volovik1996}, or the electrons localized on impurities.

The adiabatic condition (ii) is violated when the inverse relaxation time
$1/\tau$ is comparable with the minigap -- the interlevel distance
of Andreev bound states in the vortex core -- which is
$\omega_0 \sim \Delta^2/E_F$. In this case one should either solve the transport
equation or consider different limits -- different classes with zero dissipation
separated by the regions of the parameters, where the  dissipation is finite.
In the translationally invariant system these limits are determined by the
value of the parameter
$\omega_0\tau$:   in the hydrodynamic regime, $\omega_0\tau\ll 1$, the Magnus
force is very small $\rho_M\sim \rho \Delta^2/E_F^2$; in
the  collisionless regime,  $\omega_0\tau\gg 1$, the {\it ideal} value of the
Magnus force is restored, $\rho_M= \rho$. For the
$d$-wave superconductors with anisotropic gap one has even two different
parameters, coming from the semiclassical and the true (quantum) minigaps
\cite{KopninVolovik1997,Makhlin}, and thus one has 3 different regimes.

In
the  hydrodynamic regime, $\omega_0\tau\ll 1$, the essential reduction of the
Magnus force occurs due to the spectral flow of the electrons (momentum exchange
between the core electrons and the heat bath).
Formally this extreme spectral flow means the pinning of the
core electrons by the heat bath or by the crystal lattice. That is
why the effects of violation of conditions (i) and (ii) should be
similar. Formally the electrons localized in the translationally
noninvariant system have $\tau=0$, which corresponds to the
hydrodynamic regime and thus to the extreme spectral flow.

We argue that  the origin of the negligibly small Magnus
force in the SIS JJA results from
the strong violation of the translation invariance, due to which almost all
electrons are pinned within islands. It appears that the reduction
of the Magnus force is effectively the same as in SNS JJA: the
Magnus force is nonzero only due to the small asymmetry between particles and
holes \cite{MakhlinVolovik}. The origin of this coincidence is that
both the spectral flow phenomenon in SNS JJA and the smallness of the
Josephson coupling  between the islands in SIS JJA lead to the similar
effect of pinning of almost all the electrons: they do not follow the vortex
dynamics and do not contribute the Magnus force.

Quantitatively the Magnus force is determined by the linear in $\dot
\phi_i$ term of the effective action for the $i$-th island (see Eq.(97) of
\cite{Sonin}):
\begin{equation}
{1\over 2}\tilde N_i \dot \phi_i~~.
\label{TopAction}
\end{equation}
The Magnus force is proportional to $\tilde N$ according to Eq.(100) of
Ref.\cite{Sonin}. The quantity $\tilde N$ can be calculated within the BCS
theory applied to the superconductor in the
$i$-th island. The calculation of the effective BCS action shows that
the variable $\tilde N_i$, which is canonically conjugated to the phase
$\phi_i$, is not the total number $N_i$ of electrons but is
proportional the square of the gap amplitude (see eg. Ref.\cite{Otterlo}).
In a more general form the quantity $\tilde N_i$ can be written as
\cite{Volovik1997}
\begin{equation}
\tilde N_i(\Delta)= \int_0^\Delta d\Delta'
{dN_i(\Delta')\over
d\Delta'}= N_i(\Delta) - N_i(0)~~,
\label{tildeN}
\end{equation}
where $\Delta$ is again the gap amplitude and $N_i(\Delta)$ is the number of
the electrons as a function of $\Delta$ at given chemical potential, if the
Coulomb effects are neglected. The quantity
$\tilde N_i$ differs from  the actual number
$N_i=N_i(\Delta)$ of the electrons in the superconducting island by the
constant value
$ N_i(0)$. This
$ N_i(0)=N_i(\Delta=0)$  is the number of electrons in the hypothetical
normal state  with $\Delta=0$, which has the same chemical
potential  as the superconducting state, if the charging
effect is neglected (see also Ref.\cite{vanOtterlo1995}). The quantity $\tilde
N_i$ is nonzero only due to the small asymmetry between the particles and holes
and is of order
$N_i \Delta^2/E_F^2$.

Since the parameter $ N_i(0)$ is constant, it
does not influence the classical dynamic equations for JJA, which thus remain
symmetric in the sense discussed by Sonin. But this parameter $
N_i(0)$ essentially influences the Magnus force, which contains the small factor
$N_i -  N_i(0)$ instead of
$N_i$, ie $\rho_M\sim mn \Delta^2/E_F^2$.

As was noted in Ref.\cite{Otterlo} the derivation of the
"topological term" in the BCS action,  which is linear in $\dot
\phi_i$, does not depend on such details as the electronic mean free
path and thus is the same in a clean and dirty limits. This is
confirmed by the general form of the Eq.(\ref{tildeN}). That is why the
Eq.(\ref{tildeN})   can be applied to many different limiting cases, though
one should realize in which corner of the parameter space the system is. In our
case the decription in terms of the phases $\phi_i$ of islands  is to be valid,
which implies that the critical current should be small enough. The detailed
calculations are needed to understand how the Coulomb blockade can influence
the magnitude of the topological term.

Thus the effective Lagrangian for JJA should contain two types of the
"offset charges":
\begin{eqnarray}
\nonumber
L= {1\over 2} \sum_{ij}(Q_i+Q_{1i})C_{ij}^{-1}(Q_j+Q_{1j})
\\-E_J\sum_{<ij>}\cos(\phi_i-\phi_j)
~+~{1\over 2e}(Q_i+Q_{2i}) \dot
\phi_i ~.
\label{Action}
\end{eqnarray}
Here $Q_{i}=Q_{pi}-eN_i$ is the total electric charge of the island, where
 $Q_{pi}$ is the charge of the positive ionic background; $Q_{1i}$ is
the conventional offset charge (see eg. Ref.\cite{FazioOtterloSchoen});
while
$Q_{2i}$ is the "offset charge" coming from the particle-hole
asymmetry. This charge
$Q_{2i}=e N_i(0)-Q_{pi} $ is small due to the small value of the
parameter
$\Delta^2/E_F^2$. The Magnus force is proportional to the average value of
$<Q_i> + <Q_{2i}>=<Q_{2i}>-<Q_{1i}>$. While $<Q_{1i}>$ can be zero
\cite{Otterlo}, the "offset charge" $<Q_{2i}>$ coming from the particle-hole
asymmetry does not average to zero and should produce small but finite Magnus
force.

In conclusion, contrary to the Sonin's arguments \cite{Sonin} the
symmetry of the dynamic equations for the JJA does not forbid the so called
topological term in the effective action for JJA even in the classical limit.
This term leads to the finite effective Magnus force which is proportional to
the parameter of the particle-hole asymmetry.

\end{document}